\documentclass[reprint,amsmath,amssymb,superscriptaddress,aps,prb,showkeys]{revtex4-2}
\usepackage[english]{babel}
\usepackage[utf8]{inputenc}
\usepackage{graphicx}
\usepackage{fancyhdr}
\usepackage[active]{srcltx}
\usepackage{setspace}
\usepackage{float}
\usepackage{color}
\usepackage{multirow}
\usepackage{bm}
\usepackage{amsmath, amsfonts, amssymb, mathtools}	
\usepackage{latexsym}
\usepackage{soul}

\usepackage[breaklinks, hidelinks]{hyperref}
\usepackage{xcolor}
\hypersetup{
colorlinks,
linkcolor={red!50!black},
citecolor={blue!50!black},
urlcolor={blue!80!black}
}

\begin{document}

\title{Resolving High-Energy States of Interlayer Excitons in MoSe$_2$/WSe$_2$ Heterostructures}

\author{Chirag Chandrakant Palekar}
\affiliation{Institut für Physik und Astronomie, Technische Universität Berlin Hardenbergstrasse 36, 10623 Berlin, Germany}

\author{Paulo E. {Faria~Junior}}
\email{paulo@ucf.edu}
\affiliation{Department of Physics, University of Central Florida, Orlando, Florida 32816, USA}
\affiliation{Department of Electrical and Computer Engineering, University of Central Florida, Orlando, Florida 32816, USA}
\affiliation{Institute of Theoretical Physics, University of Regensburg, 93040 Regensburg, Germany}

\author{Tobias Manthei}
\affiliation{Institut für Physik und Astronomie, Technische Universität Berlin Hardenbergstrasse 36, 10623 Berlin, Germany}

\author{Maxmilian Nagel}
\affiliation{Institut für Physik und Astronomie, Technische Universität Berlin Hardenbergstrasse 36, 10623 Berlin, Germany}

\author{Bhabani Sankar Sahoo}
\affiliation{Institut für Physik und Astronomie, Technische Universität Berlin Hardenbergstrasse 36, 10623 Berlin, Germany}

\author{Shachi Machchhar}
\affiliation{Institut für Physik und Astronomie, Technische Universität Berlin Hardenbergstrasse 36, 10623 Berlin, Germany}

\author{Imad Limame}
\affiliation{Institut für Physik und Astronomie, Technische Universität Berlin Hardenbergstrasse 36, 10623 Berlin, Germany}

\author{Martin Podhorský}
\affiliation{Institut für Physik und Astronomie, Technische Universität Berlin Hardenbergstrasse 36, 10623 Berlin, Germany}

\author{Jaroslav Fabian}
\affiliation{Institute of Theoretical Physics, University of Regensburg, 93040 Regensburg, Germany}

\author{B\'arbara Rosa}
\email{barbaraltr@gmail.com}
\affiliation{Institut für Physik und Astronomie, Technische Universität Berlin Hardenbergstrasse 36, 10623 Berlin, Germany}
\affiliation{Current address: Institute of Physics “Gleb Wataghin”, State University of Campinas, 13083-859 Campinas, Brazil}

\author{Stephan Reitzenstein}
\email{stephan.reitzenstein@physik.tu-berlin.de}
\affiliation{Institut für Physik und Astronomie, Technische Universität Berlin Hardenbergstrasse 36, 10623 Berlin, Germany}


\begin{abstract}
High-energy states of interlayer excitons (IXs) in van der Waals heterostructures remain largely unexplored despite their importance for understanding many-body interactions and nonlinear optical phenomena. Here, we use photoluminescence excitation (PLE) spectroscopy to resolve a Rydberg-like series of excited IX states in MoSe$_2$/WSe$_2$ heterostructures. We observe multiple PLE resonances below the intralayer exciton energies, which we assign to the 2s-, 3s-, and 4s-like states of the IXs. These resonances are consistently observed across heterostructures with different twist angles, indicating that the high-energy state spectrum is only weakly affected by the twist angle. Wannier-exciton calculations incorporating screened Coulomb interactions reproduce the overall energy scale and qualitative trends of the measured Rydberg-like series, supporting the assignment of the observed resonances. Our findings demonstrate that PLE provides direct experimental access to the previously unexplored high-energy IX states in van der Waals heterostructures.
\end{abstract}


\maketitle


\section{Introduction} \label{intro}

Excitonic phenomena in heterostructures (HSs) composed of transition metal dichalcogenide (TMDC) monolayers (MLs) have attracted considerable attention owing to their unique optical and electronic properties. TMDC-based HSs host multiple excitonic resonances arising from their type-II band alignment, including intra- and interlayer excitons \cite{Rivera2015,Nayak2017,Jiang2021,rosa2024}. While intralayer excitons exhibit strong light--matter coupling due to their large oscillator strength and substantial binding energy, interlayer excitons (IXs) possess distinct optical properties, including electrically tunable emission energies, valley-polarised photoluminescence, and long lifetimes resulting from the spatial separation of electrons and holes \cite{Rivera2015, Rivera2018valley, jauregui2019electrical, Jiang2021, FariaJunior2023}. In addition, the twist angle and lattice mismatch between the constituent MLs give rise to a moiré superlattice potential that can localize excitons and strongly modify their optical response \cite{Nayak2017, Yu2017, Tran2019}. Since the moiré potential can be tuned through the twist angle, TMDC HSs provide a versatile platform for engineering unconventional excitonic states in two-dimensional semiconductors \cite{Palekar2024, Seyler2019Nature, Brem2020}.

The exceptionally large exciton binding energies in TMDC MLs, typically on the order of several hundreds of meVs due to reduced dielectric screening and strong quantum confinement \cite{Mak2010,Wang2018}, give rise to well-resolved Rydberg-like exciton series. Investigations of these high-energy states provide direct insight into long-range Coulomb interactions and dielectric screening effects \cite{Chernikov2014}. In contrast, high-energy IX states in TMDC HSs remain largely unexplored because of their weak oscillator strength. Although previous studies have suggested the existence of these states \cite{Barre2022Science}, direct experimental observation has remained elusive. In particular, conventional reflection contrast ($\Delta R / R$) measurements provide only limited access to these weak optical resonances, making the identification of high-energy IX states particularly challenging.

Here, we use photoluminescence excitation (PLE) spectroscopy to resolve a Rydberg-like series of excited IX states in MoSe$_2$/WSe$_2$ HSs. We investigate heterostructures with twist angles of $1^\circ$, $14^\circ$, and $56^\circ$, and observe multiple PLE resonances below the corresponding intralayer exciton energies, which we assign to the 2s-, 3s-, and 4s-like excited states of the IXs. These resonances are consistently observed across all fabricated HSs, indicating that the excited-state spectrum is only weakly affected by the twist angle. To support our experimental observations, we perform Wannier-exciton calculations for HSs on SiO$_2$, encapsulated in hBN, and suspended in air, revealing the influence of dielectric screening on the high-energy states spectrum. Together, our experimental and theoretical results establish a Rydberg-like series of high-energy IX states and demonstrate that PLE spectroscopy provides direct experimental access to these previously unexplored excitonic states.


\begin{figure*}[!ht]
	\centering\includegraphics[width=17cm]{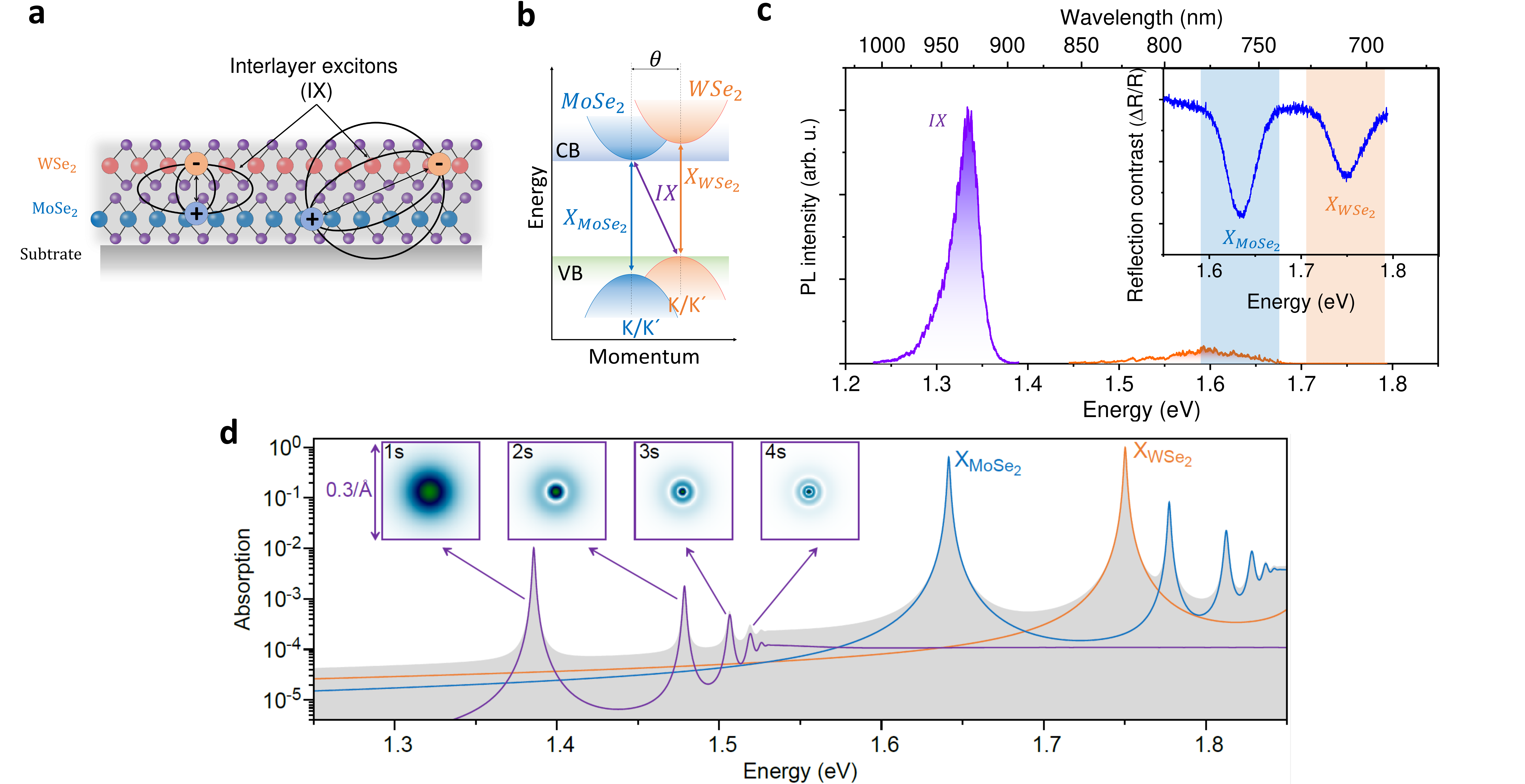} 
	\caption{\textbf{Excitons in MoSe$_2$/WSe$_2$ heterostructures.} \textbf{a)} Schematic of the in MoSe$_2$/WSe$_2$ heterostructure along with IX states. \textbf{b)} Schematic of the type-II band alignment at K/K' and the associated intra- as well as inter-layer exciton transitions. \textbf{c)} PL emission of MoSe$_2$/WSe$_2$ heterostructures measured at 4~K exhibiting pronounced emission from IXs. Inset: Reflection contrast measurement focusing on the intralayer exciton (X\textsubscript{MoSe\textsubscript{2}} and X\textsubscript{WSe\textsubscript{2}}) absorption. \textbf{d)} Normalised calculated absorption spectrum of the intra- and inter-layer excitons for an air/MoSe$_2$/WSe$_2$/SiO$_2$ configuration. The insets show the wavefunctions of the Rydberg-like IXs (1s, 2s, 3s and 4s excited states) in k-space from $-$0.15 $\textrm{\AA}^{-1}$ to 0.15 $\textrm{\AA}^{-1}$.}
	\label{Figure_1}
\end{figure*}

\section{Interlayer excitons} \label{IX in HS}

The MoSe$_2$/WSe$_2$ HSs were fabricated by conventional mechanical exfoliation \cite{Novoselov2004} followed by dry transfer \cite{Castellanos-Gomez2014}. Details of the sample preparation are provided in the Methods Section \ref{Sample Preparation}. During the processing, HSs with different twist angles were fabricated, allowing us to examine the robustness of the excitonic signatures against twist-angle variations. Further details on the determination of the twist angle are provided in the Methods Section \label{Spectroscopy Methods} and Supplementary Information (SI) Fig. S1.

Figure \ref{Figure_1}(a) schematically illustrates a MoSe$_2$/WSe$_2$ HS together with the relevant excitonic species, while Figure \ref{Figure_1}(b) shows the corresponding type-II band alignment and the associated intra- and interlayer excitonic transitions. We investigate these excitonic transitions at 4~K using photoluminescence (PL) spectroscopy under above-band excitation at 1.878~eV (660~nm). The PL spectrum of the HS region, shown in Figure \ref{Figure_1}(c), is dominated by a broad IX emission at low energy together with strongly quenched intralayer exciton emission from X$_{\mathrm{MoSe_2}}$ and X$_{\mathrm{WSe_2}}$ at higher energies. The quenching of the intralayer excitons is predominantly attributed to ultrafast charge transfer between the constituent MLs, which efficiently converts intralayer excitons into IXs. In contrast, the reflection contrast spectrum shown in the inset of Figure \ref{Figure_1}(c) clearly resolves the X$_{\mathrm{MoSe_2}}$ and X$_{\mathrm{WSe_2}}$ resonances at 1.634~eV and 1.750~eV, respectively. The measured intra- and interlayer exciton energies are consistent with previous reports \cite{Rivera2015,Hanbicki2018,Palekar2024_trilayer}.

To gain further insight into the excitonic spectrum of the MoSe$_2$/WSe$_2$ HS, we calculate its optical absorption spectrum, as shown in Figure \ref{Figure_1}(d). The calculated lowest-energy intra- and interlayer exciton resonances are in excellent agreement with the experimentally observed spectral response in Figure \ref{Figure_1}(c). More importantly, the calculations reveal a Rydberg-like series of IX states, with the corresponding momentum-space wavefunctions of the 1s, 2s, 3s, and 4s states displayed in the insets. The 1s state corresponds to the lowest-energy IX transition observed in PL, whereas the excited states progressively approach the free-carrier continuum with increasing principal quantum number. Consequently, the energy spacing between successive states decreases, while the excitonic wavefunctions become increasingly localized in momentum space and, correspondingly, more extended in real space. These trends reflect the progressively weaker binding of the high-energy IX states and highlight the challenge of experimentally probing them. Details of the calculations are provided in the Methods Section \ref{Calculations}.


\begin{figure*}[!ht]
	\centering\includegraphics[width=17cm]{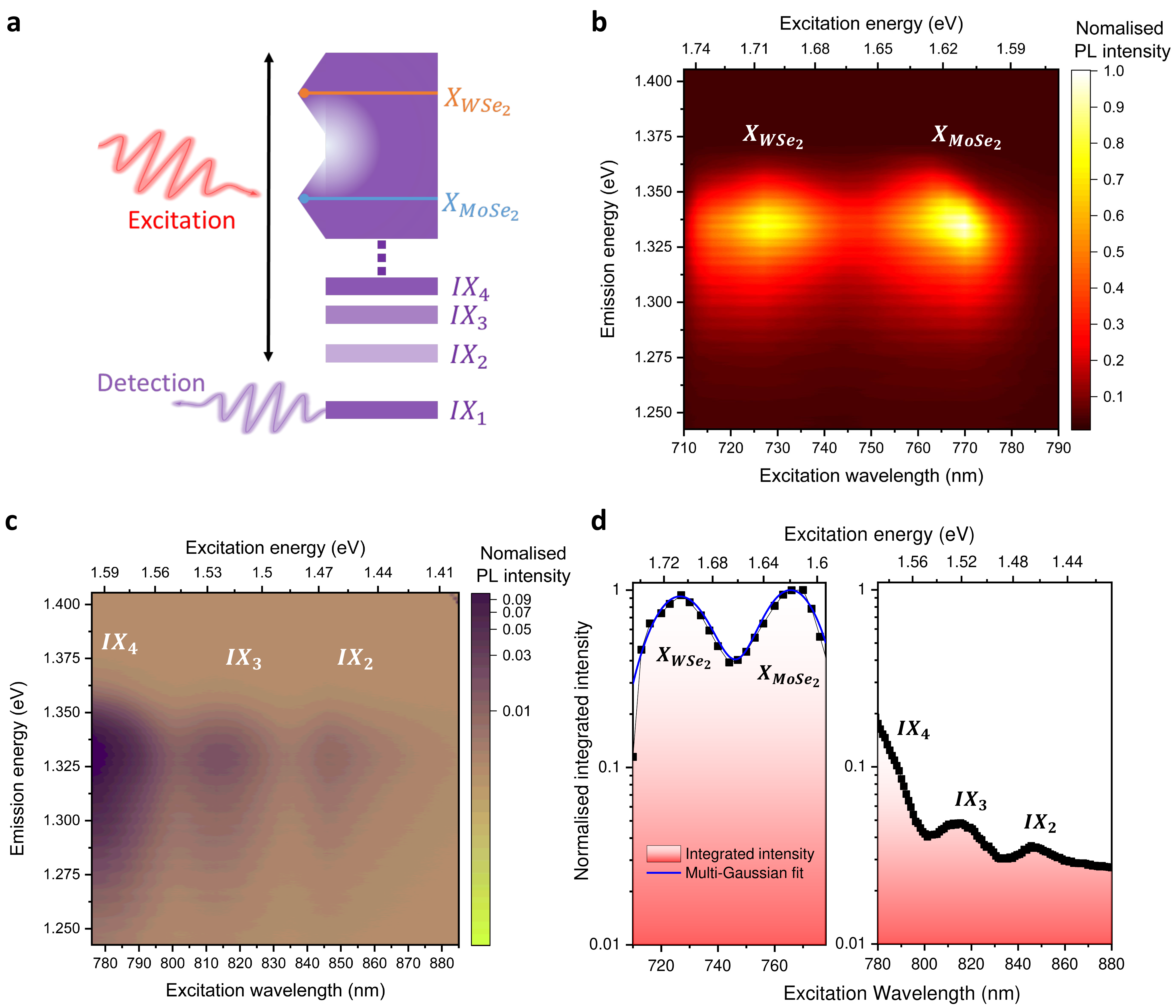} 
	\caption{\textbf{Identification of high-energy IX states through photoluminescence excitation spectroscopy.} \textbf{a)} Schematic of the excitation and detection energetics of the PLE measurements. The PLE measurements on a MoSe$_2$/WSe$_2$ HSs with a twist angle of 56$^\circ$ are shown in \textbf{b)} and \textbf{c)}. The false-color graph demonstrates IX state emission as a function of excitation energy (wavelength) under constant excitation power. The intensity is normalised in both \textbf{b)} and \textbf{c)} plots with respect to MoSe$_2$ resonance. \textbf{d)} Normalised integrated intensity (black squares) from data shown in \textbf{b)} (left panel) and \textbf{c)} (right panel), respectively as a function of the excitation wavelength exhibiting resonances associated with the IX (X\textsubscript{MoSe\textsubscript{2}} and X\textsubscript{WSe\textsubscript{2}}). The integrated intensity is fitted with a Gaussian function (blue line) in \textbf{d)} (left panel). Interestingly, \textbf{d)} (right panel) shows two additional PLE resonances corresponding to IX$_2$, IX$_3$ and IX$_4$.}
	\label{Figure_2}
\end{figure*}

\section{Signatures of high-energy interlayer exciton states} \label{PLE main}

The high-energy IX states possess even weaker oscillator strengths than the IX ground state, making them inaccessible by conventional PL spectroscopy. Consequently, probing these states through reflection contrast spectroscopy, which has been successfully employed to investigate Rydberg-like intralayer excitons in TMDC MLs \cite{Chernikov2014, fang2019control, Barre2022Science}, remains extremely challenging. In contrast, PLE spectroscopy provides a highly sensitive approach for probing weak optical resonances and has previously enabled the observation of Rydberg-like intralayer excitons in TMDC MLs \cite{Hill2015}. Beyond identifying excitonic resonances, PLE also provides insight into the relaxation pathways leading to the detected emission.

Here, we employ PLE spectroscopy to probe excited IX states in MoSe$_2$/WSe$_2$ HSs. The measurement scheme is illustrated in Figure~\ref{Figure_2}(a), where the IX emission intensity is recorded as a function of the excitation wavelength (or energy) while maintaining a constant excitation power. The resulting false-color map is shown in Figure~\ref{Figure_2}(b). As expected, the IX emission is enhanced when the excitation laser is resonant with the X$_{\mathrm{MoSe_2}}$ and X$_{\mathrm{WSe_2}}$ intralayer excitons, producing pronounced PLE resonances consistent with previous studies \cite{Seyler2019Nature, Li2020_PLE, Michl2022}.

Interestingly, IX emission persists with significantly low intensity even when the excitation energy is below the X$_{\mathrm{MoSe_2}}$ resonance. To investigate this unexpected behavior, we monitored the IX emission while tuning the excitation energy between the IX and X$_{\mathrm{MoSe_2}}$ transitions. The corresponding PLE intensity map is presented in Figure~\ref{Figure_2}(c). In addition to the intralayer exciton resonances, three distinct intensity maxima emerge below the X$_{\mathrm{MoSe_2}}$ transition. We assign these resonances to the 2s-, 3s-, and 4s-like states of the IX, hereafter denoted as IX$_2$, IX$_3$, and IX$_4$, respectively. The IX ground state (1s) is correspondingly denoted as IX$_1$. This assignment is strongly supported by the excellent agreement between the measured resonance energies and the calculated Rydberg-like IX series shown in Figure~\ref{Figure_1}(d).

To quantify these resonances, we fitted the excitation-energy-dependent integrated PL intensity using a multi-Gaussian function. The resulting fits are shown in Figure~\ref{Figure_2}(d) (right panel). From the fits, we extract PLE resonance energies of (1.706$\pm$0.002)~eV and (1.618$\pm$0.001)~eV for X$_{\mathrm{WSe_2}}$ and X$_{\mathrm{MoSe_2}}$, respectively. These resonance energies are red shifted by 43~meV and 15~meV relative to the corresponding absorption energies extracted from the reflection contrast measurements in Figure~\ref{Figure_1}(c). This difference is expected because PLE reflects not only the optical absorption associated with a given excitonic transition but also the subsequent carrier relaxation and energy transfer processes leading to IX emission \cite{Michl2022,Palekar2024}. Consequently, PLE resonances do not necessarily coincide with the corresponding absorption resonances.

Furthermore, Figure~\ref{Figure_2}(d) (left panel) demonstrates the high-energy IX states (IX$_2$, IX$_3$, and IX$_4$) through the intensity modulations observed in excitation-energy-dependent integrated PL intensity. It is evident that the intensity modulations are significantly less pronounced in comparison to intralayer exciton PLE resonances. Hence, additional evidence is obtained supporting the assignment of high-energy IX states from the excitation-energy dependence of the emission line width and energy of IX$_1$. Although the detected emission always originates from the IX$_1$ ground state, both quantities exhibit weak but distinct modulations at excitation energies corresponding to the higher-order IX resonances, as shown in SI Fig.~S2. In particular, the emission linewidth varies by approximately 5~meV, reflecting changes in the excitation and relaxation pathways rather than intrinsic modifications of the IX$_1$ state itself. Similar linewidth modulations are also observed for X$_{\mathrm{WSe_2}}$ and X$_{\mathrm{MoSe_2}}$ at their respective resonance energies. In addition, the IX$_1$ emission exhibits a small blue shift accompanied by linewidth narrowing at the resonance energies. The correlated evolution of the emission energy and linewidth provides further evidence that the observed PLE resonances originate from high-energy IX states.

\begin{figure} [!ht]
	\centering\includegraphics[width=\columnwidth] {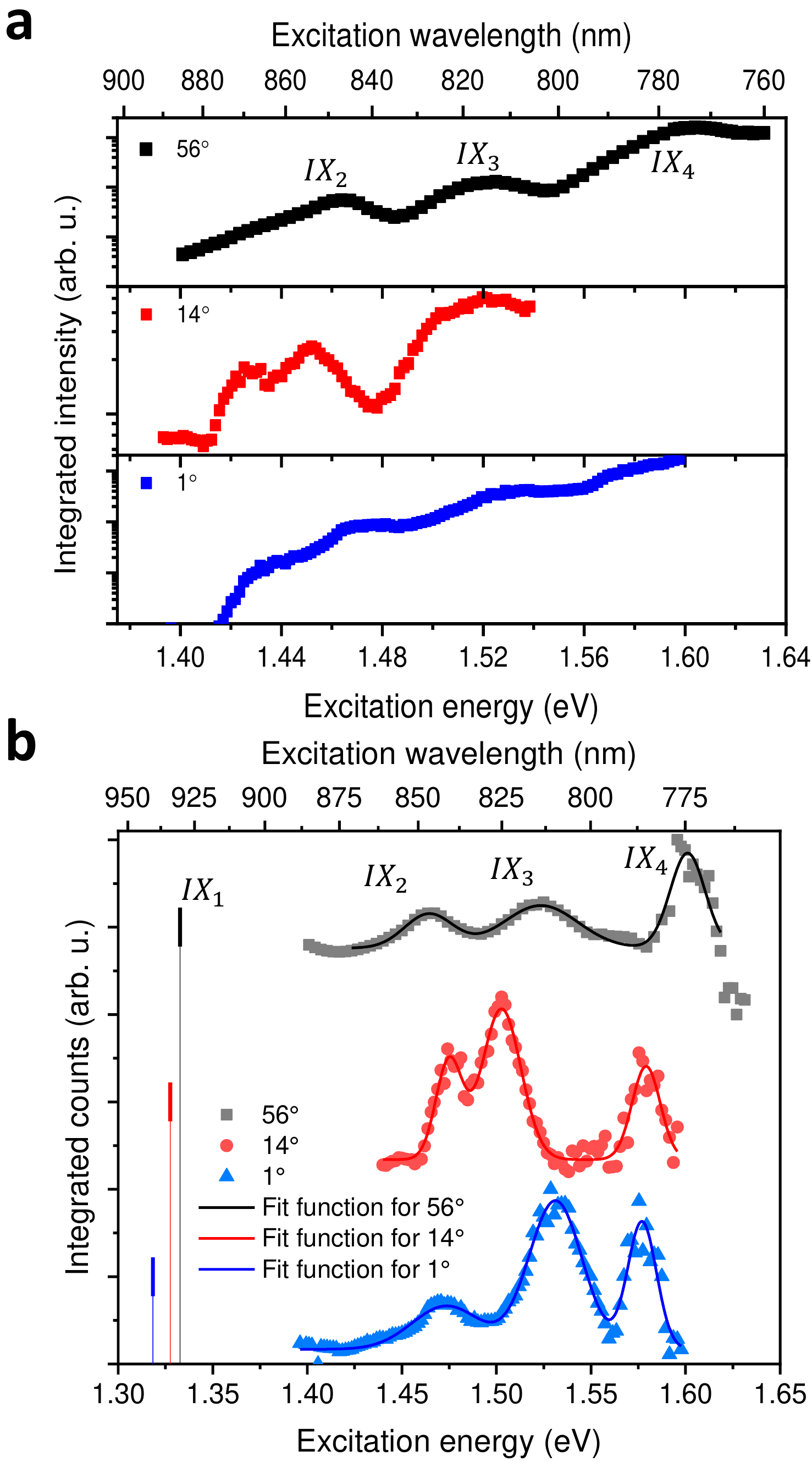} 
	\caption{\textbf{Rydberg-like high-energy states of IX states in MoSe$_2$/WSe$_2$ HSs.} \textbf{a)} Integrated intensity acquired from PLE measurements from the HSs with twist angles of 56$^\circ$ (black), 14$^\circ$ (red), and 1$^\circ$ (blue) from top to bottom, respectively. \textbf{b)} Integrated intensity resonance subtracted from exponentially decaying baseline. The resultant data is fitted with multi-Gauss function to extract the energetic separation between the states. The vertical dash lines at lower energies indicated the 1s IX state for each twist angle along with their projection on to the x axis.  
    }
	\label{Figure_3}
\end{figure}

\begin{figure*}[!ht]
	\centering\includegraphics[width=16cm] {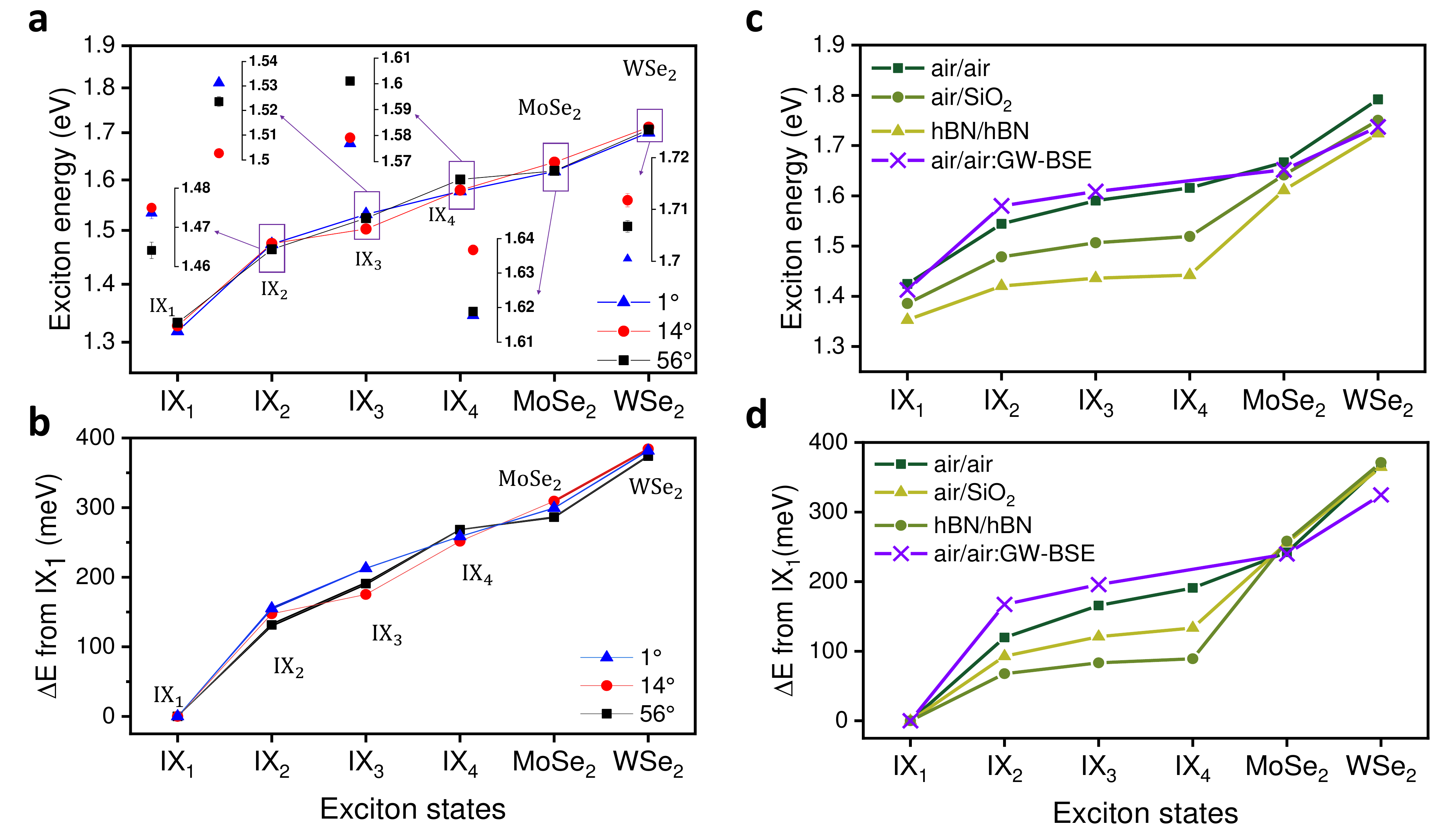} 
	\caption{\textbf{Twist-angle robustness and dielectric screening dependence of high-energy IX states.} \textbf{a)} Extracted energies of the Rydberg-like states of the IX from Figure \ref{Figure_3} \textbf{b)} for HSs with varying twist angle along with the intralayer exction PLE resonances (MoSe$_2$ and WSe$_2$). The extracted values are highlighted further with reduced energy scale in insets. \textbf{b)} Extracted energy separation ($\Delta E$) with respect to ground state of IX$_1$ for high energy IX states in HS with varying twist angle. \textbf{c)} Calculated exciton energies for different HS cases: suspended (air/air), on SiO$_2$ substrate (air/SiO$_2$), and hBN encapsulated (hBN/hBN). The dashed line corresponds to energy values of a suspended HS obtained from GW-BSE calculations of Ref.~\cite{Barre2022Science}. \textbf{d)} Calculated energy separation between the IX ground state and high-energy states for similar stacking scenarios as \textbf{b)}.
    }
	\label{Figure_4}
\end{figure*}


\section{Robust energy dependence of high-energy interlayer excitons}

The twist angle between the constituent MLs of a TMDC HS is a key parameter for tuning the properties of interlayer excitons \cite{Nayak2017,Choi2021,Palekar2024}. We therefore investigate whether the excited IX states identified by PLE spectroscopy exhibit a similar sensitivity to the twist angle. Surprisingly, despite the pronounced twist-angle dependence of the IX ground-state emission energy, the excited-state spectrum remains remarkably robust across all investigated samples.

Figure~\ref{Figure_3}(a) shows the integrated IX intensity as a function of excitation energy for MoSe$_2$/WSe$_2$ HSs with twist angles of $1^\circ$, $14^\circ$, and $56^\circ$. The analysis follows the same procedure described for Figure~\ref{Figure_2}(c). To highlight the weak PLE resonances, the integrated intensity is plotted on a logarithmic scale. Similar to the $56^\circ$ sample, clear high-energy IX resonances are observed for the $1^\circ$ and $14^\circ$ HSs, demonstrating that the excited-state ladder is present over a broad range of twist angles.

The baseline-subtracted PLE spectra together with the corresponding multi-Gaussian fits are shown in Figure~\ref{Figure_3}(b). The vertical lines indicate the centroid energies of the IX$_{1s}$ emission for each sample. From the fits, we extract the resonance energies of the excited IX states. For the $56^\circ$ HS, the IX$_{2s}$, IX$_{3s}$, and IX$_{4s}$ resonances are located at (1.464 $\pm$ 0.002)~eV, (1.523 $\pm$ 0.002)~eV, and (1.601 $\pm$ 0.001) eV, respectively.

The extracted resonance energies for all three HSs are summarized in Figure~\ref{Figure_4}(a). While the PLE resonances of X$_{\mathrm{MoSe_2}}$ and X$_{\mathrm{WSe_2}}$ remain nearly unchanged, except for moderate blue shifts of the $14^\circ$ sample, the IX$_{1s}$ emission exhibits the expected blue shift with increasing twist angle \cite{Nayak2017}. In contrast, the high-energy IX resonances display only small energy variations of 9~meV, 28~meV, and 24~meV for the IX$_{2s}$, IX$_{3s}$, and IX$_{4s}$ states, respectively. These observations indicate that the excited-state spectrum is considerably less sensitive to the twist angle than the IX ground state.

To remove the influence of the twist-angle-dependent IX$_{1s}$ energy, we analyze the energy separation $\Delta E = E_n-E_1$ between the excited states and the IX ground state, as shown in Figure~\ref{Figure_4}(b). As expected for a Rydberg-like series, the spacing between successive states decreases with increasing principal quantum number. Nevertheless, the energy scaling clearly deviates from the hydrogenic $1/n^2$ behavior (see SI Fig.~S3), reflecting the non-hydrogenic electron--hole interaction characteristic of atomically thin semiconductors \cite{Chernikov2014,Hill2015,biswas2023rydberg}.

The weak dependence of the excited-state spectrum on twist angle is somewhat unexpected given the strong influence of twist angle on many interlayer exciton properties. This observation suggests that several competing mechanisms determine the energies of the excited IX states. Besides twist-angle-induced modifications of the electronic structure and moiré potential, hybridization with intralayer excitons \cite{rosa2024}, dielectric screening, and structural disorder can all contribute to the observed resonance energies. In particular, local variations of the interlayer separation, interface roughness, trapped contaminants, and strain introduce spatially varying Coulomb interactions and dielectric screening, effects that are expected to be particularly important for the spatially extended high-energy IX states \cite{Raja2019,Plankl2021}.

To further elucidate the origin of the excited-state spectrum, we calculate the exciton energies for different dielectric environments using the Wannier-exciton approach. Figure~\ref{Figure_4}(c) presents the calculated exciton energies for suspended HSs (air/air), HSs on SiO$_2$ (air/SiO$_2$), and fully encapsulated structures (hBN/hBN). The calculations reveal a much stronger dependence on dielectric screening than on twist angle. As the dielectric screening increases, all IX states shift toward the quasiparticle continuum owing to the reduction of the electron--hole Coulomb interaction. This effect is particularly pronounced for the excited IX states because their larger spatial extent makes them more sensitive to the surrounding dielectric environment.

The effect is highlighted in Figure~\ref{Figure_4}(d), where the energies are referenced to the IX$_{1s}$ state. Our calculations are in great agreement with previous GW-BSE results for suspended HSs \cite{Barre2022Science}, demonstrating that the Wannier-exciton model accurately captures the excited IX spectrum. Increasing the dielectric screening leads to a pronounced compression of the excited-state ladder, substantially reducing the energy separation between IX$_{1s}$ and the higher-order states. The stronger dielectric response of the high-energy IX states reflects their reduced binding energies and increased Bohr radii.

Taken together, the experimental and theoretical results indicate that the excited-state spectrum is remarkably robust against twist-angle variations, while remaining highly sensitive to the dielectric environment. This behavior suggests that dielectric screening, together with disorder and exciton hybridization, plays a more important role than twist angle in determining the energies of the high-energy IX states. The overall agreement between experiment and theory further supports the assignment of the observed PLE resonances to the Rydberg-like excited states of interlayer excitons.
 

\section{Conclusions} \label{Conclusion}

In summary, we investigated the high-energy states of interlayer excitons in MoSe$_2$/WSe$_2$ heterostructures using PLE spectroscopy. We identified multiple PLE resonances below the intralayer exciton energies and assigned them to the IX$_{2s}$, IX$_{3s}$, and IX$_{4s}$ excited states, providing experimental evidence for a Rydberg-like series of interlayer excitons. The observed resonances persist across heterostructures with different twist angles, indicating that the excited-state spectrum is only weakly affected by the twist angle.

Wannier-exciton calculations incorporating screened Coulomb interactions reproduce the overall energy scale and qualitative trends of the measured Rydberg-like series and reveal a substantially stronger dependence on the dielectric environment than on the twist angle. Together, the experimental and theoretical results identify dielectric screening as a key parameter governing the excited-state spectrum of interlayer excitons. Our work demonstrates that PLE spectroscopy provides direct experimental access to the previously unexplored excited-state spectrum of interlayer excitons, opening new opportunities to investigate their fundamental properties and role in van der Waals heterostructures.


\section*{Acknowledgements}

C.C.P., B.S.S., B.R., and S.R. acknowledge the financial support of the Deutsche Forschungsgemeinschaft (DFG, German Research Foundation) via SPP 2244 (Project No. 443416027). S.M. and S.R. acknowledge financial support of the Berlin Senate via Berlin Quantum. P.E.F.J. and J.F. acknowledge the financial support of the DFG via SFB 1277 (Project-ID 314695032, projects B07 and B11), SPP 2244 (Project No. 443416183), and of the European Union Horizon 2020 Research and Innovation Program under Contract No. 881603 (Graphene Flagship). P.E.F.J. acknowledges the computational resources of the Advanced Research Computing Center of the University of Central Florida.


\section*{Methods}
\label{Methods}

\subsection*{A. TMDC Sample Preparation}\label{Sample Preparation}

The multilayered TMDC samples were fabricated using mechanical exfoliation \cite{Novoselov2004} and the dry-transfer method \cite{Castellanos-Gomez2014}. For exfoliation of the MLs, commercially available crystals of MoSe$_2$ and WSe$_2$ are used along with Netto blue tape and Polydimethylsiloxane (PDMS). Bulk crystals of MoSe$_2$ and WSe$_2$ are mechanically exfoliated on the PDMS gel strip with the help of blue tape. Next, the individual MLs are identified by optical contrast microscopy and PL spectroscopy at room temperature. Single crystalline MLs of MoSe$_2$ and WSe$_2$ are selected by observing the formed straight edges, at 60$^\circ$ or 120$^\circ$, as an indication of crystal axes. For the HS assembaly fabrication, the MoSe$_2$ ML was transferred onto a SiO$_2$/SiN DBR deposited on Si substrate. During the stacking process, the WSe$_2$ ML was aligned to the edge of MoSe$_2$ ML, addressing the required twist angles before transferring it on the the MoSe$_2$ ML. Following a similar sample preparation method, a multiple samples were prepared.

\subsection*{B. Spectroscopy Methods}\label{Spectroscopy Methods}

\textbf{Second harmonic generation:} The twist angle between the constituent layers of the samples was determined by performing polarization resolved second harmonic generation (SHG) measurements \cite{xubaka2013}. The TMDC ML was excited with linearly polarised picosecond mode-locked laser with wavelength of 1313~nm. Then, the SHG intensity (at 656~nm) as function of excitation laser polarization is recorded from the ML transferred on the substrate. The characteristic intensity maxima show a six-fold symmetry, and each maximum indicates the armchair direction on the hexagonal crystal lattice of the associated TMDC ML. Comparing the SHG response from the constituent ML of the heterostructure system, the twist angle between the ML can be determined. To differentiate between R- or H-type stacking, we measured the SHG response of the HS regions. Reduction in intensity of SHG signal from the HS region with respect to MLs indicates H-type stacking. Similarly, R-type stacking results in higher SHG signal than H-type stacking as it leads to restored inversion symmetry \cite{Hsu2014SecondHG}. The experimental uncertainty in the range of 1$^\circ$ to 3$^\circ$.

\textbf{Photoluminescence spectroscopy:} PL spectroscopy was used to measure the excitonic emission from the MLs and HS systems at room temperature to identify MLs as well as at low temperature to study the IX emission from HS regions. For ML identification, continues wave excitation laser was used with wavelength at 532~nm. For low temperature measurements, a wavelength tunable picosecond mode lock pulsed laser was used as source of excitation to perform the PLE measurements. For the PLE measurements, the IX emission intensity was recorded by tuning the laser from 705~nm to 885~nm while keeping the excitation power constant over the tuning range. Low temperature measurements were conducted at 4~K using close-cycle cryostat equipped with a high numerical aperture (NA = 0.81) objective lens. 

\subsection*{C. Exciton Calculations}\label{Calculations}

The Wannier-exciton calculations were performed using the effective BSE equation~\cite{Rohlfing2000PRB,Zollner2020PRB} considering parabolic bands for electrons and holes and a screened Coulomb potential for intra- and inter-layer excitons within the HS successfully employed in Ref.~\cite{Ovesen2019CommPhys}. The parameters used in the calculations (effective masses, dielectric constants, and layer thicknesses) are taken from Ref.~\cite{Ovesen2019CommPhys}. The BSE is solved on a 2D $k$-grid from -0.4 to 0.4 $\textrm{\AA}^{-1}$ in $k_x$ and $k_y$ directions with total discretization of $241 \times 241$ points. To improve convergence, the Coulomb potential (intra- and inter-layer) is averaged around each $k$-point in a square region of $-\Delta k /2$ to $\Delta k /2$ discretized with $241 \times 241$ points~\cite{Zollner2020PRB}. The calculated absorption spectra includes a Lorentzian broadening with energy dependent full width at half-maximum following Ref.~\cite{Zollner2020PRB}. For the broadening parameters, we assume $\Gamma_1 = 1.5 \; \text{meV}$ and $\Gamma_2 = \Gamma_3 = 20 \; \text{meV}$ and $E_0$ 20~meV above the onset of the continuum (single-particle band gap). To account for the band gap reduction due to the dielectric screening\cite{Cho2018PRB}, we consider a reduction of 140~meV for air/SiO$_2$ and 230~meV for hBN/hBN. The band gaps of MoSe$_2$ and WSe$_2$ in the HS are considered as 1.995~eV and 2.067~eV, respectively, taken from Ref.~\cite{Ovesen2019CommPhys}.


\bibliography{biblio}

\end{document}


\preprint{APS/123-QED}

\title{Supplementary Information: Resolving High-Energy States of Interlayer Excitons in MoSe$_2$/WSe$_2$ Heterostructures}

\author{Chirag Chandrakant Palekar}
\affiliation{Institut für Physik und Astronomie, Technische Universität Berlin Hardenbergstrasse 36, 10623 Berlin, Germany}

\author{Paulo E. {Faria~Junior}}
\email{paulo@ucf.edu}
\affiliation{Department of Physics, University of Central Florida, Orlando, Florida 32816, USA}
\affiliation{Department of Electrical and Computer Engineering, University of Central Florida, Orlando, Florida 32816, USA}
\affiliation{Institute of Theoretical Physics, University of Regensburg, 93040 Regensburg, Germany}

\author{Tobias Manthai}
\affiliation{Institut für Physik und Astronomie, Technische Universität Berlin Hardenbergstrasse 36, 10623 Berlin, Germany}

\author{Maxmilian Nagel}
\affiliation{Institut für Physik und Astronomie, Technische Universität Berlin Hardenbergstrasse 36, 10623 Berlin, Germany}

\author{Bhabani Sankar Sahoo}
\affiliation{Institut für Physik und Astronomie, Technische Universität Berlin Hardenbergstrasse 36, 10623 Berlin, Germany}

\author{Shachi Machchhar}
\affiliation{Institut für Physik und Astronomie, Technische Universität Berlin Hardenbergstrasse 36, 10623 Berlin, Germany}

\author{Imad Limame}
\affiliation{Institut für Physik und Astronomie, Technische Universität Berlin Hardenbergstrasse 36, 10623 Berlin, Germany}

\author{Martin Podhorský}
\affiliation{Institut für Physik und Astronomie, Technische Universität Berlin Hardenbergstrasse 36, 10623 Berlin, Germany}

\author{Jaroslav Fabian}
\affiliation{Institute of Theoretical Physics, University of Regensburg, 93040 Regensburg, Germany}

\author{B\'arbara Rosa}
\email{barbaraltr@gmail.com}
\affiliation{Institut für Physik und Astronomie, Technische Universität Berlin Hardenbergstrasse 36, 10623 Berlin, Germany}
\affiliation{Current address: Institute of Physics “Gleb Wataghin”, State University of Campinas, 13083-859 Campinas, Brazil}

\author{Stephan Reitzenstein}
\email{stephan.reitzenstein@physik.tu-berlin.de}
\affiliation{Institut für Physik und Astronomie, Technische Universität Berlin Hardenbergstrasse 36, 10623 Berlin, Germany}

\maketitle

\setcounter{table}{0}
\renewcommand{\tablename}{Table S}
\renewcommand{\thetable}{\arabic{table}}

\setcounter{figure}{0}
\renewcommand{\figurename}{Figure S}
\renewcommand{\thefigure}{\arabic{figure}}


\section{Twist-angle determination based on SHG measurements}

\begin{figure*}[!ht]
	\centering\includegraphics[width=18cm] {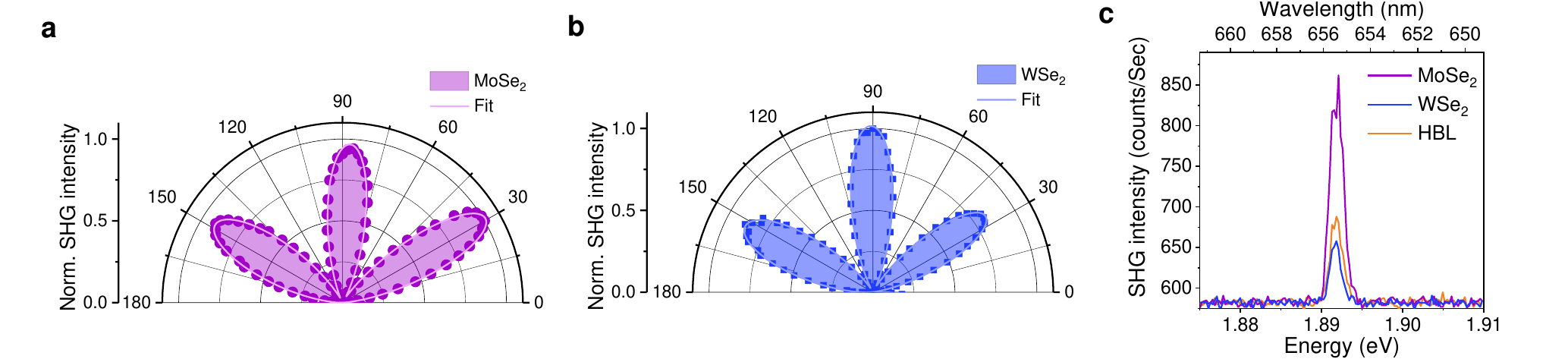} 
	\caption{\textbf{ Twist-angle and stacking determination by SHG.}  The polar plot of SHG response from the MLs, \textbf{(a)} MoSe$_2$ and \textbf{(b)} WSe$_2$, measured under pulsed laser excitation. \textbf{(c)} Comparison between the SHG signals acquired from both MLs and HS region. The reduced SHG intensity from the HS region compared to both MLs corresponds to the 3H type stacking. Fitting yields a twist angle of $56^\circ \pm 0.5^\circ$.}
	\label{SI-fig1}
\end{figure*}

Polarization-resolved second-harmonic generation (SHG) measurements were performed to determine the twist angle and stacking configuration of the WSe$_2$/MoSe$_2$ HS, as SHG is highly sensitive to crystal orientation and symmetry \cite{Hsu2014SecondHG,Malard2013,Michl2022}. The ML and HS regions were excited with a linearly polarized pulsed laser at 1313~nm, and the SHG signal was detected at 656~nm. The excitation polarization was rotated from $0^\circ$ to $180^\circ$, while the polarization-dependent SHG intensity was recorded for both constituent MLs [Figure S~\ref{SI-fig1}(a,b)].

The measured SHG patterns were fitted using the sum of three Gaussian functions,

\begin{equation}
\label{eq:SHG_intensity}
I_{\mathrm{SHG}}=\sum_{i=1}^{3}\left(y_0+\frac{A}{\omega\sqrt{\pi/2}}
\exp\left[-\frac{2(x-x_c^{i})^2}{\omega^2}\right]\right) \, ,
\end{equation}
\\
where $A$ is the peak area, $\omega$ is the peak width, $x_c^{i}$ denotes the peak positions, and $y_0$ is a constant offset. The extracted peak positions are summarized in Table S~\ref{tab:SHG_peaks}.

\begin{table}[h]
\centering
\renewcommand{\arraystretch}{1.3}
\begin{tabular}{|c|c|c|c|c|}
\hline
Material & Peak 1 & Peak 2 & Peak 3 & Twist angle \\
\hline
WSe$_2$ & $34.2^\circ$ & $90.7^\circ$ & $152.7^\circ$ & $3.9^\circ \pm 0.5^\circ$ \\
MoSe$_2$ & $30.6^\circ$ & $87.3^\circ$ & $148.4^\circ$ &  \\
\hline
\end{tabular}
\caption{Peak positions extracted from polarization-resolved SHG measurements.}
\label{tab:SHG_peaks}
\end{table}

The angular offset between the SHG patterns of the two monolayers yields a relative twist angle of $3.9^\circ\pm0.5^\circ$. To determine the stacking configuration, the SHG intensity from the HS region was compared with that of the individual monolayers under identical excitation conditions [Figure S~\ref{SI-fig1}(c)]. The pronounced suppression of the SHG signal in the HS indicates a centrosymmetric 2H-like stacking configuration. Consequently, the effective twist angle of the heterostructure is obtained as $60^\circ-3.9^\circ=56.1^\circ\pm0.5^\circ$.


\section{Signatures of interlayer exciton higher order states}

\begin{figure}[!ht]
	\centering\includegraphics[width=17cm]{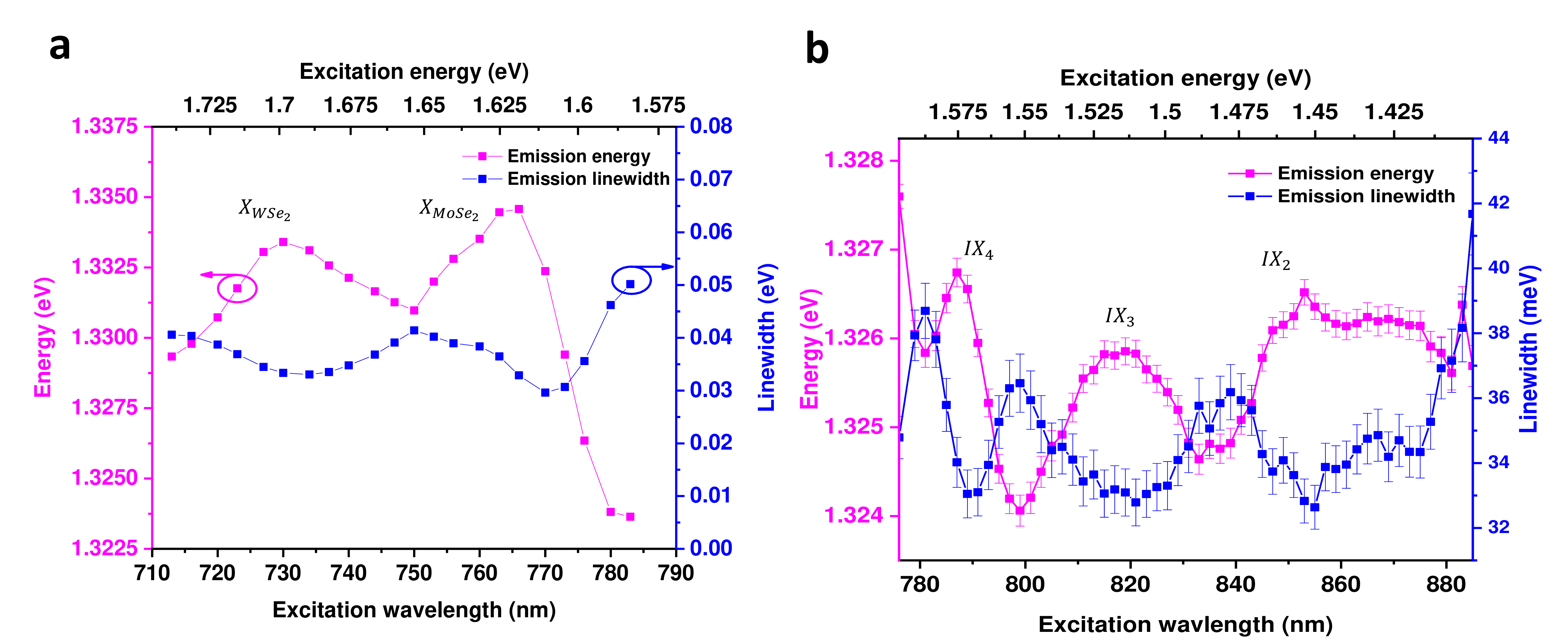} 
	\caption{\textbf{Emission energy and linewidth of IX$_1$. a)} shows the extracted emission energy and linewidth modulations of IX$_{1s}$ at spectral position of X\textsubscript{WSe\textsubscript{2}} and X\textsubscript{MoSe\textsubscript{2}}. Similarly, \textbf{b)} shows the extracted emission energy and linewidth exhibiting signatures of high-energy states such IX$_{2s}$, IX$_{3s}$ and IX$_{4s}$.}
	\label{SI-fig2}
\end{figure}

The IX emission as function of excitation wavelength is fitted with the Lorentz function to account for inhomogeneous boarding and to extract the emission linewidth. Figure S \ref{SI-fig2} reveals that the IX emission PL intensity and linewidth exhibit a correlation, which provide signatures of the resonance. The linewidth of IX reduces considerably around the IX higher order states, along with an increase in the emission intensity.

\newpage
\section{Non-hydrogenic nature of high-energy IX states}

\begin{figure}[!ht]
	\centering\includegraphics[width=10cm]{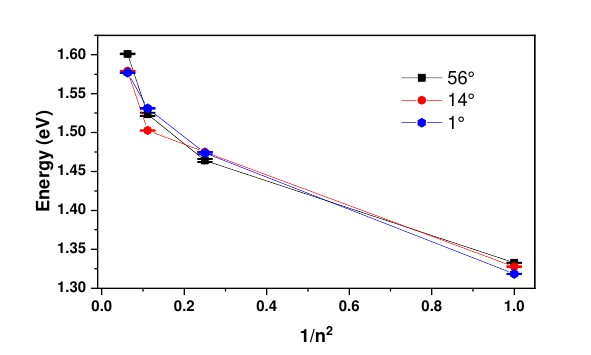} 
	\caption{\textbf{Comparison to hydrogenic dependence.} Extracted PLE resonance energies of Rydberg like high-energy IX states from all HS samples with twist angle 56°, 14° and 1° as function of 1/n$^2$.}
	\label{SI-fig3}
\end{figure}


\bibliography{biblio}